\pdfoutput=1
\newcommand*{\ATLASLATEXPATH}{latex/}
\documentclass[PAPER, atlasdraft=true, texlive=2016, UKenglish, cernpreprint]{\ATLASLATEXPATH atlasdoc}
\usepackage[backend=bibtex,block=none]{\ATLASLATEXPATH atlaspackage}
\usepackage{\ATLASLATEXPATH atlasbiblatex}
\usepackage{\ATLASLATEXPATH atlascontribute}
\usepackage{\ATLASLATEXPATH atlasphysics}

\graphicspath{{logos/}{figures/}}

\usepackage{EXOT-2017-02-PAPER-defs}

\AtlasTitle{Search for $W' \rightarrow tb$ decays in the hadronic final state using $pp$ collisions at $\sqrt{s}=13~\TeV$ with the ATLAS detector}

\AtlasAbstract{%
A search for $W'$-boson production in the $W' \rightarrow t\bar{b} \rightarrow q\bar{q}' b\bar{b}$ decay channel is presented using \lumi\ of 13~\TeV\ proton--proton collision data collected by the ATLAS detector at the Large Hadron Collider in 2015 and 2016. The search is interpreted in terms of both a left-handed and a right-handed chiral \wp\ boson within the mass range 1--5 TeV. Identification of the hadronically decaying top quark is performed using jet substructure tagging techniques based on a shower deconstruction algorithm. No significant deviation from the Standard Model prediction is observed and the results are expressed as upper limits on the $W' \rightarrow t\bar{b}$ production cross-section times branching ratio as a function of the $W'$-boson mass. These limits exclude $W'$ bosons with right-handed couplings with masses below 3.0 TeV and $W'$ bosons with left-handed couplings with masses below 2.9 TeV, at the 95$\%$ confidence level.
}

\author{The ATLAS Collaboration}

\AtlasRefCode{GROUP-2017-02}

\PreprintIdNumber{CERN-EP-2017-340}

\AtlasJournal{Phys.\ Lett.\ B.}
\AtlasJournalRef{\PLB 781 (2018) 327}
\AtlasDOI{10.1016/j.physletb.2018.03.036}

\hypersetup{pdftitle={ATLAS document},pdfauthor={The ATLAS Collaboration}}

\begin{document}

\maketitle

\clearpage
\section{Introduction}
\label{sec:into}
\begin{figure}[!h!tpb]
  \centering
\scalebox{.8}{
  \includegraphics[trim={0 0 0.8cm 0},clip,width=0.6\textwidth]{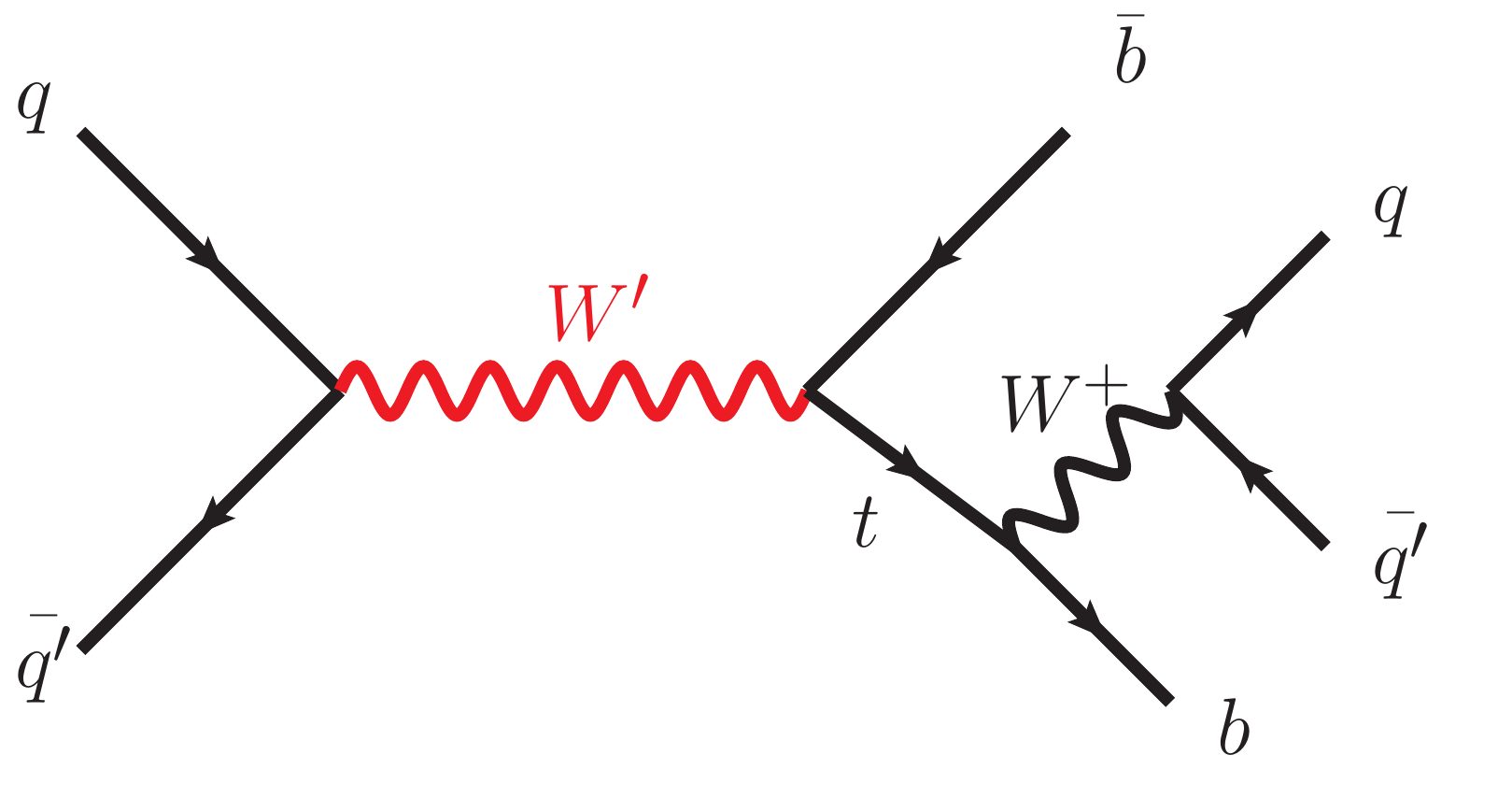}
}
  \caption{Feynman diagram for \wp-boson production with decay into $t\bar{b}$ and a hadronically decaying top quark.}
  \label{fig:feynman}
\end{figure}

Several theories beyond the Standard Model (SM) involve enhanced symmetries that predict new gauge bosons,
usually called \wp\ or $Z'$ bosons. The \wp\ boson is the mediator of a new charged vector current that
can be massive enough to decay into a top quark and a $b$-quark (as in Figure~\ref{fig:feynman}). Many models such as those with extra dimensions~\cite{DIENES199947}, strong dynamics~\cite{PhysRevD.13.974,PhysRevD.20.2619,DIMOPOULOS1979237,EICHTEN1980125}, composite Higgs~\cite{Muller:1996dj}, or the Little Higgs~\cite{1126-6708-2002-07-034,PhysRevD.59.075003} predict new vector charged-current interactions, some with preferential couplings to quarks or third-generation particles~\cite{Kaluza-Klein,Technicolor,left-right-model,Muller:1996dj,HILL1995483}. Due to the large mass of the top quark, its interactions decouple from the rest of the phenomenology in many theories beyond the SM. An effective Lagrangian is used to capture the relevant phenomenology of the Sequential Standard Model (SSM)~\cite{Altarelli1989} \wp\ $\rightarrow$ $t\bar{b}$ signal~\cite{ZackSullivan,Duffty:2012rf}, which has the same coupling strength to fermions as the SM $W$ boson but higher mass. 

Searches for a \wp\ boson decaying into $t\bar{b}$,\footnote{
For simplicity, the notation ``$t\bar{b}$'' is used to denote the final state for both $W'^{+} \to t\bar{b}$ and  $W'^{-} \to \bar{t}b$ decays.} classified as either leptonic or hadronic according to the decay products of the $W$ boson originating from the top quark, were performed at the Tevatron~\cite{PhysRevLett.103.041801,Abazov2011145} and the Large Hadron Collider (LHC) in final states that include leptons~\cite{Sirunyan:2017vkm,ATLASWprime8TeV,CMSWprime,CMSWprime8TeV} or that are fully hadronic~\cite{EXOT-2013-14}. The specific search for a \wp\ boson decaying into $t\bar{b}$ allows for a right-handed \wp\ boson (\wpr) in models in which the right-handed neutrino's mass is assumed to be much higher than that of the \wp\ boson ($m_{\nu_R}$ > $m_{W'}$), which the leptonic decay mode cannot access. In such a model, the branching ratio for a \wpr\ boson decaying into $t\bar{b}$ is $\mathcal{O}$(10$\%$) higher relative to that for a \wpl\ boson decaying into $t\bar{b}$ since a \wpl\ boson can also decay to a lepton and neutrino. Limits on a SSM left-handed \wp\ boson (\wpl) decaying into a lepton and a neutrino have been set previously~\cite{ATLASWplep,CMSWplep}. Previous searches in the all-hadronic final state exclude \wpr\ bosons with masses up to 2~TeV, set at the 95$\%$ confidence level (CL) using 20.3 $fb^{-1}$ of $pp$ collision data at a centre-of-mass energy ($\sqrt{s}$) of 8~\TeV~\cite{EXOT-2013-14}. A recent search by the CMS Collaboration in the lepton+jets final state excludes \wpr-boson masses up to 3.6 TeV using 35.9 $fb^{-1}$ of $pp$ collision data collected at $\sqrt{s} = 13~\TeV$~\cite{Sirunyan:2017vkm}.

\newcommand{\AtlasCoordFootnote}{%
ATLAS uses a right-handed coordinate system with its origin at the nominal interaction point (IP)
in the centre of the detector and the $z$-axis along the beam pipe.
The $x$-axis points from the IP to the centre of the LHC ring,
and the $y$-axis points upwards.
Cylindrical coordinates $(r,\phi)$ are used in the transverse plane,
$\phi$ being the azimuthal angle around the $z$-axis.
The pseudorapidity is defined in terms of the polar angle $\theta$ as $\eta = -\ln \tan(\theta/2)$.
Angular separation is measured in units of $\Delta R \equiv \sqrt{(\Delta\eta)^{2} + (\Delta\phi)^{2}}$, where $\Delta\eta$ and $\Delta\phi$ are the separations in $\eta$ and $\phi$. Momentum in the transverse plane is denoted by \pt.}

This analysis searches for a \wp\ boson decaying into $t\bar{b}$ with a mass in the range of 1--5 TeV, in the invariant mass spectrum of the top quark and bottom quark ($m_{tb}$) reconstructed in the fully hadronic channel. This includes a \wpr\ boson that is not kinematically allowed to decay into a lepton and neutrino and a \wpl\ boson that can decay into quarks or leptons. The large \wp\ mass results in a top quark and a $b$-quark that have high transverse momentum (\pt).\footnote{\AtlasCoordFootnote}
The decay products of the top quark become more collimated as the top-quark \pt increases, and their showers partially overlap~\cite{PERF-2015-04}. This high-\pt\ topology is referred to as \textit{boosted}. 
The boosted top-quark decay is reconstructed as a single jet. The shower deconstruction (SD) algorithm~\cite{PhysRevD.84.074002,PhysRevD.87.054012} is employed to select, or tag, jets from boosted top-quark decays. A signal would be reconstructed as a localised excess in the $m_{tb}$ distribution rising above the smoothly falling background originating mostly from jets created by the strong interaction described by quantum chromodynamics (QCD). This analysis represents an improvement on the previous ATLAS analysis in this channel~\cite{EXOT-2013-14} due to a higher centre-of-mass energy, higher integrated luminosity, and better top-tagging techniques, understanding of systematic uncertainties, and statistical treatment.

\section{ATLAS detector}
\label{sec:atlas}
The ATLAS detector~\cite{PERF-2007-01} at the LHC covers almost the entire solid angle around the collision point. Charged particles in the pseudorapidity range |$\eta$| < 2.5 are reconstructed with the inner detector (ID), which consists of several layers of semiconductor detectors (pixel and strip) and a straw-tube transition-radiation tracker, the latter covering |$\eta$| < 2.0. The high-granularity silicon pixel detector provides four measurements per track; the closest layer to the interaction point is known as the insertable B-layer (IBL)~\cite{ATLAS-TDR-19}. The IBL was added in 2014 and provides high-resolution hits at small radius to improve the tracking performance. The ID is immersed in a 2 T magnetic field provided by a superconducting solenoid. The solenoid is surrounded by electromagnetic and hadronic calorimeters, and a muon spectrometer incorporating three large superconducting toroid magnet systems.
The calorimeter system covers the pseudorapidity range |$\eta$| < 4.9. Electromagnetic calorimetry is performed with barrel and endcap high-granularity lead/liquid-argon (LAr) electromagnetic calorimeters, within the region |$\eta$| < 3.2. There is an additional thin LAr presampler covering |$\eta$| < 1.8, to correct for energy loss in material upstream of the calorimeters. For |$\eta$| < 2.5, the LAr calorimeters are divided into three layers in depth. Hadronic calorimetry is performed with a steel/scintillator-tile calorimeter, segmented into three barrel structures within |$\eta$| < 1.7, and two copper/LAr hadronic endcap calorimeters, which cover the region 1.5 < |$\eta$| < 3.2. The forward solid angle up to |$\eta$| = 4.9 is covered by copper/LAr and tungsten/LAr calorimeter modules, which are optimised for energy measurements of electrons/photons and hadrons, respectively.
The muon spectrometer (MS) comprises separate trigger and high-precision tracking chambers that measure the deflection of muons in a magnetic field generated by superconducting air-core toroids. The ATLAS detector uses a tiered trigger system to select interesting events. The first level is implemented in custom electronics and reduces the event rate from the LHC crossing frequency of 40 MHz to a design value of 100 kHz. The second level is implemented in software running on a general-purpose processor farm which processes the events and reduces the rate of recorded events to $\sim$1 kHz~\cite{Aaboud:2016leb}.

\section{Data and simulation samples}
\label{sec:datasim}

This analysis uses data from proton--proton ($pp$) collisions at $\sqrt{s} = 13~\TeV$ collected with the ATLAS detector in 2015 and 2016 that satisfy a number of criteria to ensure that the ATLAS detector was in good operating condition. The amount of data used in this analysis corresponds to an integrated luminosity of \lumi. The average number of $pp$ interactions delivered per LHC bunch crossing was 23.7.

Monte Carlo (MC) event generators were used to simulate signal and background events. Signal events were generated at leading order (LO) in QCD by \textsc{MadGraph5$\_$aMC@NLO v2.2.3}~\cite{MadGraph5two}, using a chiral \wp-boson model in which the coupling strength of the \wp\ boson to the right- and left-handed fermions are the same as those of the SM $W$ boson to left-handed fermions. The \wpl\ boson can decay into all left-handed fermions, but the \wpr\ boson can decay only into right-handed quarks as the right-handed neutrino is assumed to be more massive than the \wpr\ boson. \textsc{MadGraph} was used to simulate the top-quark and $W$-boson decays, taking spin correlations into account. \textsc{Pythia v8.186}~\cite{Pythia8} was used for the modelling of the parton shower, fragmentation and the underlying event. The \textsc{NNPDF23LO} parton distributions function (PDF) set~\cite{NNPDF} and the A14 set of tuned parameters~\cite{ATL-PHYS-PUB-2014-021} were used for the event generation. All simulated samples were rescaled to next-to-leading-order (NLO) calculations using NLO/LO $K$-factors ranging from 1.3 to 1.4, depending on the mass and handedness of the \wp\ boson, calculated with \textsc{Ztop}~\cite{Duffty:2012rf}. The width of the \textsc{MadGraph} simulated \wp\ boson is set to the NLO \textsc{Ztop} width calculation, $\mathcal{O}$(3$\%$) of its mass. Signal samples with gauge-boson masses between 1 and 3 TeV were generated in 250 GeV steps, and between 3 and 5 TeV in 500 GeV steps. 

The dominant SM background process is multi-jet production. In order to reduce the dependence on the modelling of the simulation a data-driven method is implemented as described in Section~\ref{subsec:evtsel}. Corrections in this method are estimated using QCD dijet simulation produced at LO by {\textsc{Pythia v8.186}}. Uncertainties in this method are obtained using simulated QCD dijet events produced at LO by {\textsc{Herwig++ v2.7.1}}~\cite{Bahr:2008pv} and {\textsc{Sherpa v2.1.1}}~\cite{sherpa}, and at NLO by {\textsc{Powheg-Box v2}}~\cite{powheg1,powheg2} with either {\textsc{Pythia8}} or {\textsc{Herwig+Jimmy}}~\cite{Corcella:2000bw} for the parton shower, fragmentation and the underlying event simulation (referred to as {\textsc{Powheg+Pythia}} and {\textsc{Powheg+Herwig}}, respectively). Vector bosons ($W/Z$) produced in association with jets are included in the data-driven approach. These processes are expected to contribute less than 1$\%$ of the multi-jet background. This $W/Z$+jets prediction is checked using events simulated with the \textsc{Sherpa v2.2.1}~\cite{sherpa} generator and the CT10 PDF set~\cite{CT10}.

 Top-quark pair production is an important background with an inclusive cross-section of $\sigma_{\ttbar}=832 ^{+46}_{-51}$~pb for a top-quark mass of $172.5~\gev$ as obtained from calculations accurate to next-to-next-to-leading order and next-to-next-to-leading logarithms (NNLO+NNLL) in QCD with \textsc{Top++2.0}~\cite{Cacciari:2011hy, Beneke:2011mq, Baernreuther:2012ws, Czakon:2013goa, Czakon:2012pz, Czakon:2012zr, Czakon:2011xx}. Simulated top-quark pair processes were produced using the NLO \textsc{Powheg-Box v2} generator with the CT10 PDF. The parton shower, fragmentation and the underlying event were added using \textsc{Pythia} v6.42~\cite{pythia6} with the Perugia 2012 set of tuned parameters~\cite{PerugiaTunes}. To increase the number of simulated events at high mass, samples were produced binned in \ttbar mass. Interference and background contributions from the SM $s$-channel single-top process are found to be negligible and are not considered further in this analysis.

The generation of the simulated event samples includes the effect of multiple $pp$ interactions per bunch crossing, as well as the effect on the detector response due to interactions from bunch crossings before or after the one containing the hard interaction. For all \textsc{MadGraph}, \textsc{Powheg}, \textsc{Pythia} and \textsc{Herwig} samples, the EvtGen v1.2.0 program~\cite{LANGE2001152} was used for the bottom and charm hadron decays. The simulated samples were passed through the \textsc{Geant4}-based ATLAS detector simulation~\cite{geant1,geant2} and were reconstructed with the same algorithms as the data events.

\section{Event reconstruction and shower deconstruction}
\label{sec:Objsel}
\subsection{Event reconstruction}
\label{subsec:evtreco}
This analysis relies on the reconstruction and identification of jets initiated by the top- and bottom-quark daughters of the \wp\ boson. Jets are built from topologically related energy depositions in the calorimeters with the anti-$k_{t}$ algorithm~\cite{Cacciari:2008gp} using the FastJet package~\cite{Cacciari2012}. Two radius parameters are used for jet reconstruction: a small radius (small-$R$) of 0.4 and a large radius (large-$R$) of 1.0. The momenta of both the small-$R$ and large-$R$ jets are corrected for energy losses in passive material and for the non-compensating response of the calorimeter~\cite{ATL-PHYS-PUB-2015-015}. Small-$R$ jets are also corrected for the average additional energy due to pile-up interactions~\cite{PERF-2016-06}. Energy depositions from pile-up are removed from large-$R$ jets using the trimming algorithm~\cite{Krohn:2009th}: the constituents of the large-$R$ jet are reclustered using the $k_{t}$ jet algorithm~\cite{CATANI1993187,PhysRevD.48.3160} with $R$ = 0.2. Constituent jets contributing less than 5$\%$ of the large-$R$ jet's \pt are removed. The remaining energy depositions are used to calculate the trimmed-jet kinematics and substructure properties. In order to improve on the angular resolution of the calorimeter, the mass of a large-$R$ jet is computed using a combination of calorimeter and tracking information~\cite{ATLAS-CONF-2016-035}.

Small-$R$ jets are used to identify the jets compatible with originating from a \B-quark created either directly from the \wp\ boson or from the top-quark decay. Only small-$R$ jets with \pt\ > 25 GeV and |$\eta$| < 2.5 (in order to be within the coverage of the ID) are considered in this analysis. Additional \pt\ requirements are applied to enhance the sensitivity of the search (see Section~\ref{subsec:evtsel}). To reduce the number of small-$R$ jets originating from pile-up interactions, a likelihood discriminant, based on track and vertex information, is used to determine whether the primary vertex\footnote{Collision vertices are formed from tracks with \pt\ > 400 MeV. If an event contains more than one vertex candidate, the one with the highest $\sum p_{\textrm{T}}^2$ of its associated tracks is selected as the primary vertex.} is the origin of the charged-particle tracks associated with a jet candidate and rejects jets originating from pile-up interactions~\cite{PERF-2014-03}. This is done only for small-$R$ jets with \pt\ < 60 GeV and $|\eta|$ < 2.4. Small-$R$ jets which originate from \B-quarks are identified using a multivariate \B-tagging algorithm~\cite{PERF-2012-04,ATL-PHYS-PUB-2016-012}. Several observables, such as those based on the long lifetime of \B-hadrons and the \B- to $c$-hadron decay topology, are used as algorithm inputs to discriminate between \B-jets, $c$-jets and other jets. The \B-tagging requirement corresponding to an efficiency of 77$\%$ to identify \B-jets with \pt\ > 20 GeV, as determined from a sample of simulated $t\bar{t}$ events, is found to be optimal for the statistical significance of this search. This 77$\%$ working point (WP) provides rejection factors against light-flavour/gluon jets and $c$-jets of 134 and 6 respectively~\cite{ATL-PHYS-PUB-2015-022,ATL-PHYS-PUB-2016-012}. Jets identified this way are referred to as \B-tagged jets. Since the \B-tagging factors are measured in a different \pt\ region, an uncertainty is assigned to the extrapolation of the measurement to the high \pt\ region of interest.

Events with reconstructed electrons~\cite{PERF-2016-01} or muons~\cite{PERF-2015-10} are vetoed in order to ensure statistical independence of this analysis from analyses using the leptonic decay of the $W$ boson from the top quark~\cite{ATLASWprime8TeV}. Electrons and muons with transverse momenta above 25 GeV and selected with criteria similar to those used in Ref.~\cite{TOPQ-2016-01} are considered for this veto.

\subsection{Boosted-top identification using shower deconstruction}
\label{subsec:sdtop}
The SD algorithm can be used to identify the jets compatible with the hadronic decay of a $W/Z$ boson, Higgs boson, or a top quark as well as to discriminate between quark- and gluon-initiated jets. In this analysis, an SD-algorithm-based tagger (SD tagger) is used to identify jets originating from the top quark. The SD tagger calculates likelihoods that a given large-$R$ jet originates from a hadronic top-quark decay or from a high-momentum light quark or gluon. The constituents of the trimmed large-$R$ jet are used to build exclusive subjets~\cite{Cacciari2012}, and the four-momenta of these subjets serve as inputs to the SD algorithm. These subjets are used as substitutes for individual quarks and gluons originating from the hard scatter. A likelihood weight is calculated for each possible shower history that can lead to the observed subjet configuration. This step is analogous to running a parton shower MC generator in reverse, where emission and decay probabilities at each vertex, colour connections, and kinematic requirements are considered. For each shower history, the assigned weight is proportional to the probability that the assumed initial particle generates the final configuration, taking into account the SM amplitude for the underlying hard process and the Sudakov form factors for the parton shower. A variable called $\chi_{\textrm{SD}}$ is defined as the ratio of the sum of the signal-hypothesis weights to the sum of the background-hypothesis weights. For a set $\{p_{i}^{k}\}$ of $N$ observed subjet four-momenta, where $i \in [1,N]$, the value of $\chi_{\textrm{SD}}$ is given by:

\begin{equation*}
\chi_{\textrm{SD}}(\{p_{i}^{k}\}) = \frac{\sum_{\text{perm}} P(\{p_{i}^{k}\}|\text{top-quark jet})}{\sum_{\text{perm}} P(\{p_{i}^{k}\}|\text{gluon/light-quark jet})},
\end{equation*}

where $P(\{p_{i}^{k}\}|\text{top-quark jet})$ is built using the weights for the hypothesis that a signal process leads to the observed subjet configuration $\{p_{i}^{k}\}$ and $P(\{p_{i}^{k}\}|\text{gluon/light-quark jet})$ is built using the weights for the hypothesis that a background process leads to the observed subjet configuration. The $\sum_{\text{perm}}$ notation represents the sum over all the shower histories in which signal processes lead to the subjet configuration. The large-$R$ jet is tagged as a top-quark jet if $\chi_{\textrm{SD}}$ is larger than a given value, which is adjusted to achieve the desired tagging efficiency. There is an internal mechanism in the SD algorithm to suppress pile-up contributions to the jets, through the application of additional weights in the likelihood ratio, which contain the probability that a subset of the subjets did not originate from the hard interaction but from pile-up~\cite{Cacciari:2008gn}. 

The SD algorithm selects events that are kinematically compatible with a hadronic top-quark decay. The following requirements are made to optimise the algorithm to achieve a balance between good top-quark jet signal selection efficiency and rejection of gluon/light-quark jet backgrounds: the large-$R$ jet has at least three subjets; two or more subjets must have a combined invariant mass in a 60.3--100.3 GeV window centred on the $W$-boson mass; and at least one more subjet can be added to obtain a total mass in a 132--212 GeV window centred on the top-quark mass. 

The SD tagger was optimised for this analysis so that it is more efficient for top-quark jet signal selection and gluon/light-quark jet background rejection for \pt\ > 800 GeV compared to the version of the SD tagger first studied by the ATLAS Collaboration~\cite{PERF-2015-04}. This is done by building subjets obtained by using an exclusive $k_{t}$ algorithm~\cite{Cacciari2012}. First, the $k_{t}$ algorithm with $R$ = 1.0 is run over the large-$R$ jet constituents and the $k_{t}$ reclustering is stopped if the splitting scale~\cite{PhysRevLett.100.242001} is larger than 15~\gev. Once the $k_{t}$ reclustering is stopped the reclustered protojets are used as subjets. The choice of a 15~\gev\ requirement is based on the expected discrimination between signal and background events. The six highest \pt\ subjets are used as inputs to the SD algorithm. This reduces the computation time needed for the calculation of $\chi_{\textrm{SD}}$, which grows exponentially with the subjet multiplicity, without loss of background rejection power. 

\begin{figure}[!h!tpb]
  \centering
  \includegraphics[width=0.9\textwidth,trim={0 2cm 0 0},clip]{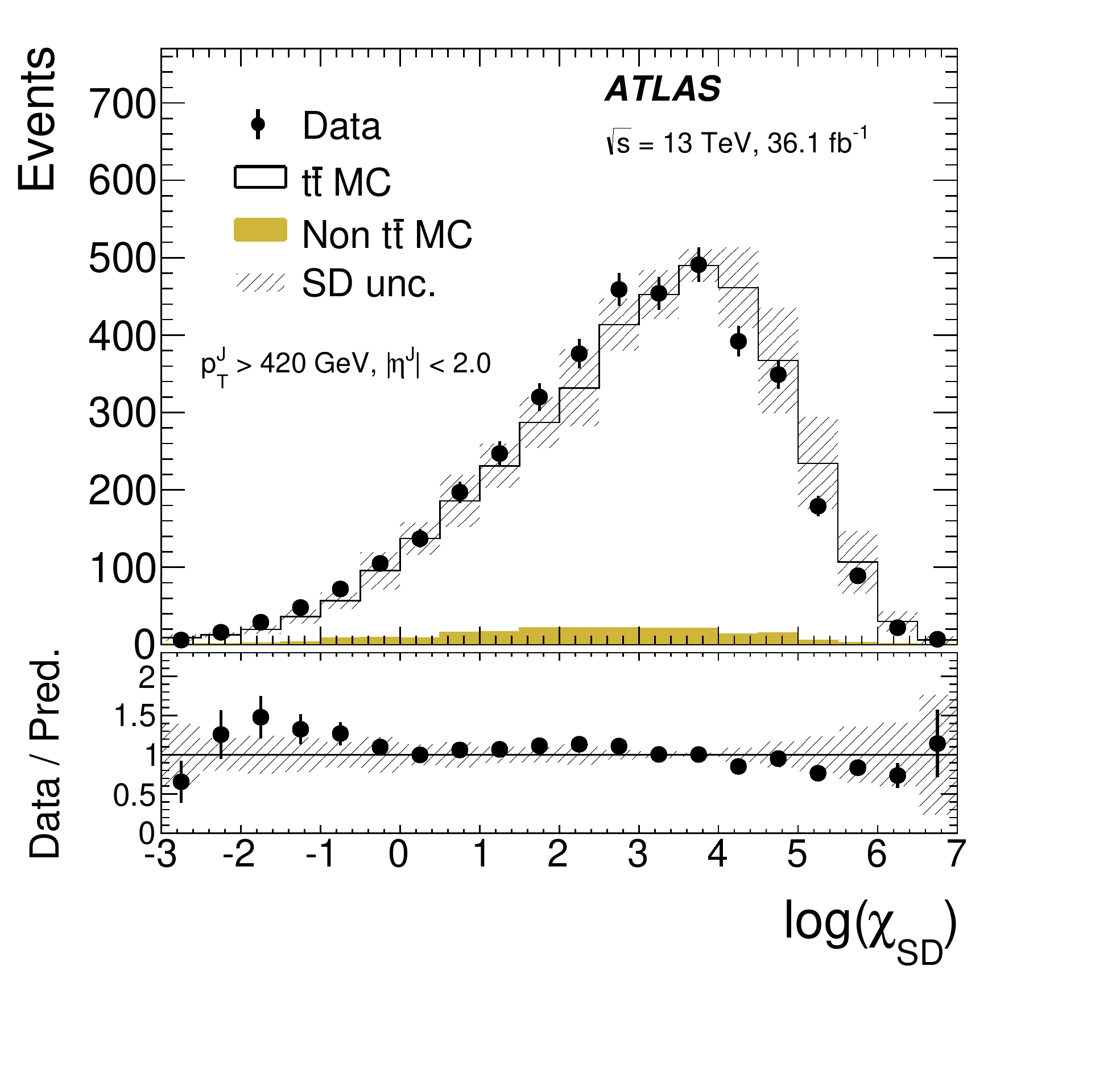}
  \caption{Comparison of the $\log(\chi_{\textrm{SD}})$ distribution between data (dots), $t\bar{t}$ MC events (line histogram) and background MC events (solid histogram) in samples with an enriched contribution from hadronically decaying top quarks using selection criteria similar to those in Ref.~\cite{PERF-2015-04}. The hatched band shows the effect of the SD uncertainty described in Section~\ref{sec:systunc}.}
  \label{fig:SD_logchi}
\end{figure}

The signal efficiency WP of the SD tagger is set by applying a selection on the logarithm of $\chi_{\textrm{SD}}$. The 50$\%$ and 80$\%$ signal efficiency WPs are used in this analysis (see Section~\ref{subsec:evtsel}). The background rejection for the 50$\%$ (80$\%$) signal efficiency WP is 80 (25) for a jet \pt\ of 0.45~TeV and 30 (10) for a \pt\ of 1.3~TeV. The $\log(\chi_{\textrm{SD}})$ variable is studied using samples enriched in hadronically decaying top quarks by selecting $t\bar{t}$ events where one top quark decays hadronically and the other into lepton+jets for events with large-$R$ jet \pt\ ($p_{\textrm{T}}^{\textrm{J}}$) > 420 GeV and $|\eta|$ ($|\eta^{\textrm{J}}|$) < 2.0. To obtain a top-quark-enriched sample, events are selected with two \B-tagged jets and either an electron or muon using criteria similar to those used in Ref.~\cite{PERF-2015-04}. The data are found to be consistent with simulation in the $\log(\chi_{\textrm{SD}})$ distribution within the SD tagger uncertainty, described in Section~\ref{sec:systunc}, as shown in Figure~\ref{fig:SD_logchi}.

\section{Event selection and background estimation}
\label{subsec:evtsel}
An initial selection of events is made at the trigger level by requiring at least one small-$R$ jet~\cite{Aaboud:2016leb} with \pt\ larger than 380 GeV. To ensure that the analysis is performed in the fully efficient regime of the trigger, the \pt\ values of the large-$R$ and small-$R$ jets, used to identify the top- and \B-quark daughters from the \wp-boson decay, are required to be larger than 420~\GeV. Candidate events must have at least one primary vertex.

The top-quark jet candidate is selected from the large-$R$ jets satisfying the requirements defined in Section~\ref{sec:Objsel}. The large-$R$ jet with the largest value of $m_j + 0.15 \times m_J$, where $m_j$ is the mass of the highest-\pt\ small-$R$ jet with minimum \pt\ > 25~\GeV\ with $\Delta R$ < 1.0 of the large-$R$ jet and $m_J$ is the mass of the large-$R$ jet, is selected as the top-quark jet candidate. This combination enhances the fraction of events where the selected large-$R$ jet is associated with the top quark, since $m_j$ is less affected by final-state radiation effects, which are important at high \pt~\cite{ATL-PHYS-PUB-2016-013}. The highest-\pt\ small-$R$ jet with \pt\ > 420 GeV and $\Delta R$ > 2.0 from the top-quark jet candidate is chosen as the \B-quark jet candidate in the event. The top- and \B-tagging criteria are applied to the selected top- and \B-quark jet candidates after rejecting events in which the \B-quark jet candidate has |$\eta$| > 1.2. This improves the signal sensitivity at high $m_{tb}$ since the high-\pt\ \B-quark jets from the \wp-boson decay tend to be more central (smaller |$\eta$|) than the jets from the multi-jet background. A summary of the top- and \B-quark jet candidate selection is shown in Table~\ref{evtsel_summary}. 

\begin{table}[t]
\caption{Summary of the top-quark jet candidate and \B-quark jet candidate selections before categorisation of events into signal and control regions. The selections are defined in Sections~\ref{sec:Objsel} and \ref{subsec:evtsel}. The events satisfying these criteria are grouped into the categories and regions described in Figure~\ref{qcd:fig1}.}
\begin{center}
\scalebox{0.9}{%
\hspace{-0.8cm}\begin{tabular}{c | c}
\hline
\hline
\multicolumn{2}{c}{ \small\textbf{Event reconstruction and selection}}  \\
\hline
Large-$R$ jet ($J$) & $p_{\textrm{T}}^J$ > 420 GeV, |$\eta$| < 2.0 \\
Small-$R$ jet ($j$) & $p_{\textrm{T}}^j$\ > 25 GeV, |$\eta$| < 2.5 \\
Top-quark jet candidate ($J_{\textrm{top}}^{\textrm{cand}}$) & jet $J$ with highest $m_{j}$ + 0.15 $\times$ $m_{J}$\\
\B-quark jet candidate ($j_{b}^{\textrm{cand}}$) & highest-\pt\ jet $j$ with $p_{\textrm{T}}^{j}$\ > 420 GeV, \\ & $\Delta R(J_{\textrm{top}}^{\textrm{cand}},j)$ > 2.0\\
\hline
Lepton veto & zero leptons with \pt\ > 25 GeV, |$\eta$| < 2.5 \\
\B-quark jet candidate $\eta$ & zero $j_{b}^{\textrm{cand}}$ with |$\eta$| > 1.2\\
\hline
0 \B-tag in & zero \B-tagged jets $j$ with $\Delta R(J_{\textrm{top}}^{\textrm{cand}},j)$ < 1.0\\
1 \B-tag in & exactly one \B-tagged jet $j$ with $\Delta R(J_{\textrm{top}}^{\textrm{cand}},j)$ < 1.0\\
\hline
\end{tabular}}
\label{evtsel_summary}
\end{center}
\end{table}

Events are divided into two categories: the ``1 \B-tag in'' category and the ``0 \B-tag in'' category. For the ``1 \B-tag in'' category, exactly one \B-tagged small-$R$ jet with $\pt > 25$~\GeV\ with $\Delta R$ < 1.0 from the top-quark jet candidate is required, while for the ``0 \B-tag in'' category, it is required that there be zero \B-tagged small-$R$ jets with $\pt > 25$~\GeV\ within the large-$R$ jet. 

\begin{figure}[htp]
\begin{center}
\hspace{0.0cm}\subfloat{\includegraphics[width=0.45\textwidth]{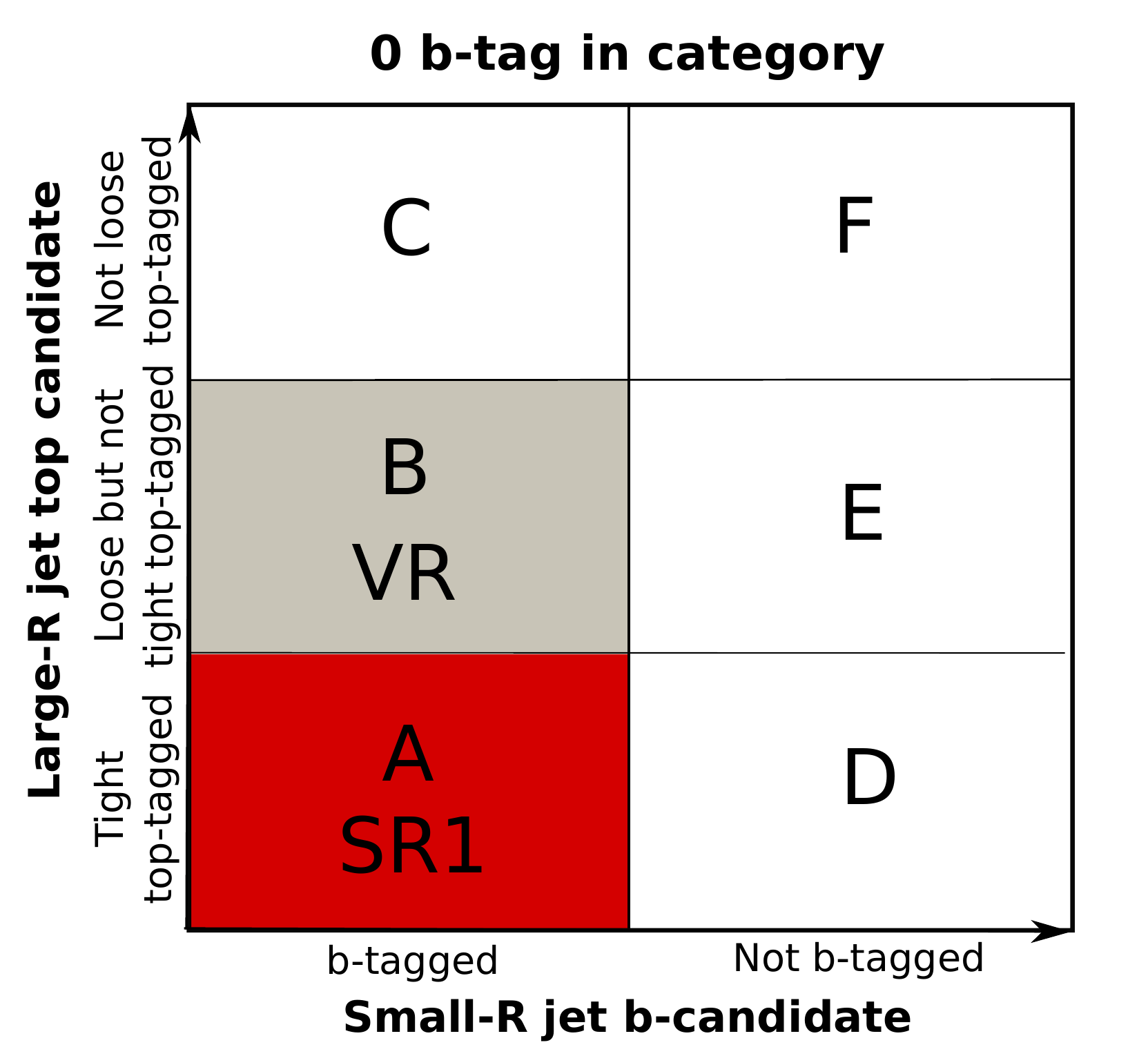}}
\hspace{0.0cm}\subfloat{\includegraphics[width=0.45\textwidth]{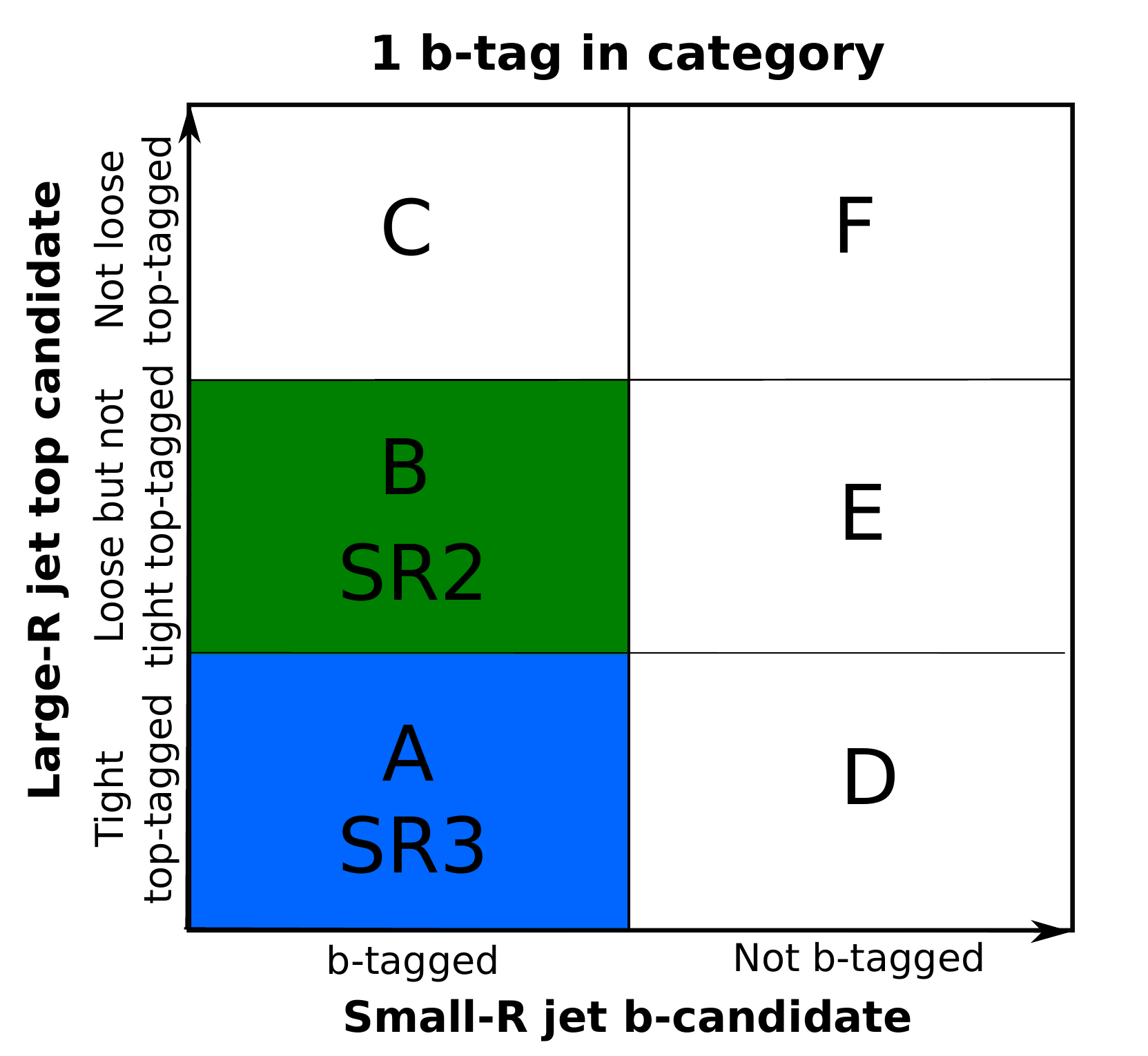}}
\caption{Illustration of the 2D sideband method showing the two-dimensional plane of the large-$R$ jet substructure variables vs the small-$R$ jet \B-tagging information used to estimate the background yield in regions A (B), from the observed yield in the three control regions C, F, D (C, F, E) for the (left) ``0 \B-tag in'' and (right) ``1 \B-tag in'' categories. The top- and \B-tagging criteria are applied after rejecting events in which the \B-quark jet candidate has |$\eta$| $>$ 1.2.}
\label{qcd:fig1}
\end{center}
\end{figure}

The binning of the $m_{tb}$ distribution is chosen to balance the sensitivity coming from the different signal and background distribution shapes against the diminishing statistical sensitivity of the data at high $m_{tb}$. Requirements are imposed on the expected number of background events per bin and the bin width is adapted to a resolution function that represents the width of the reconstructed mass peak for each studied \wp-boson signal sample. For each $m_{tb}$ bin and in each of the ``\B-tag in'' categories, the data sample is divided into six regions by using top-tagging and \B-tagging criteria, which are described in Figure~\ref{qcd:fig1}. The ``not loose top-tagged'' regions consist of events where the selected top-quark jet candidate fails to meet the loose top-tagged (80$\%$ WP) identification criteria, the ``loose-but-not-tight top-tagged'' regions consist of events where the selected top-quark jet candidate satisfies the loose top-tagged identification criteria but not the tight top-tagged criteria (50$\%$ WP) and the ``tight top-tagged'' regions consist of events where the selected top-quark jet candidate satisfies the tight top-tagged criteria. The signal regions are constructed from events in which the selected small-$R$ jet \B-candidate is \B-tagged: signal region SR1 consists of events classified as ``tight top-tagged, 0 \B-tag in'', signal region SR2 consists of events classified as ``loose-but-not-tight top-tagged, 1 \B-tag in'' and signal region SR3 consists of events classified as ``tight  top-tagged, 1 \B-tag in''. A validation region (VR), with negligible signal contamination, is defined to test the performance of the data-driven method of estimating the multi-jet + $W/Z$+jets background. This region consists of events where the \B-candidate is \B-tagged, and classified as ``loose-but-not-tight top-tagged, 0 \B-tag in''. The prediction is found to be in agreement with data within uncertainties. 

The \wp-boson signal selection efficiency, for masses below 2.5~TeV, is higher in SR2 and SR3 than in SR1, due to the requirement of zero \B-tagged jets with $\Delta R$ < 1.0 of the large-$R$ jet (``0 \B-tag in'' category) in SR1, making the topology less like the signal in SR1. For masses above 2.5~TeV, the signal efficiency is higher in SR1 than in SR2 and SR3 for the same reason: the \B-tagging efficiency, decreasing with \pt, affects SR2 and SR3 more due to the requirement of the additional \B-tagged jet. Thus, the addition of the ``0 \B-tag in'' category improves the signal sensitivity at large \wp-boson masses. The \wp-boson signal event selection efficiency is about 10$\%$ at low mass, decreasing to about 7$\%$ at high mass. The difference between the \wpr-boson and \wpl-boson signal selection efficiencies depends on the signal region and the efficiency is on average $\sim$10$\%$ higher for \wpr-boson signal samples. The difference in efficiency between \wpr-boson and \wpl-boson signals comes from a difference in angular separation between the $W$ boson and the \B-quark from the top-quark decay due to the different \wp-boson handedness, leading to a difference in the overall top-tagging efficiency.
 For instance, the 3 TeV \wpr-boson signal sample has a selection efficiency of 2.9$\%$ in SR1, 2.5$\%$ in SR2 and 2.4$\%$ in SR3, while the 3 TeV \wpl-boson signal sample has a selection efficiency of 2.7$\%$ in SR1, 2.3$\%$ in SR2 and 2.3$\%$ in SR3.

The dominant background from multi-jet production is estimated directly from data using a six-region ``2D sideband'' method that predicts both the shape and normalisation of the $m_{tb}$ distribution. These regions are shown in Figure~\ref{qcd:fig1}.

The amount of multi-jet + $W/Z$+jets background in the signal regions and in the VR is estimated bin-by-bin in the $m_{tb}$ distribution using the observed number of events in the control regions after subtracting the contribution from $t\bar{t}$ events:
\begin{center}
$N_{\mathrm{A}}^{\textrm{bkg}} = R_{\mathrm{A}}^{\mathrm{corr}}\cdot\frac{(N^{\mathrm{data}}_{\mathrm{C}}-N^{t\bar{t}}_{\mathrm{C}})\cdot(N^{\mathrm{data}}_{\mathrm{D}}-N^{t\bar{t}}_{\mathrm{D}})}{N^{\mathrm{data}}_{\mathrm{F}}-N^{t\bar{t}}_{\mathrm{F}}}$ and $N_{\mathrm{B}}^{\textrm{bkg}} = R_{\mathrm{B}}^{\mathrm{corr}}\cdot\frac{(N^{\mathrm{data}}_{\mathrm{C}}-N^{t\bar{t}}_{\mathrm{C}})\cdot(N^{\mathrm{data}}_{\mathrm{E}}-N^{t\bar{t}}_{\mathrm{E}})}{N^{\mathrm{data}}_{\mathrm{F}}-N^{t\bar{t}}_{\mathrm{F}}}$,
\end{center}
where ``bkg'' stands for multi-jet background and $W/Z$+jets background, $N_{\mathrm{A}}^{\textrm{bkg}}$ and $N_{\mathrm{B}}^{\textrm{bkg}}$ are the numbers of multi-jet + $W/Z$+jet background events in regions A and B estimated using this method; $N_{k}^{\mathrm{data}}$ and $N_{k}^{t\bar{t}}$ ($k$ = C, D, E, F) are the numbers of observed events and the expected number of $t\bar{t}$ events in each region, respectively. The correlation between the top- and \B-tagging variables ($R^{\mathrm{corr}}$) is evaluated using five simulated QCD dijet samples as:
\begin{center}
$R_{\mathrm{A}}^{\mathrm{corr}} = \frac{N_{\mathrm{A}}^{\textrm{dijet MC}} \cdot N_{\mathrm{F}}^{\textrm{dijet MC}}}{N_{\mathrm{C}}^{\textrm{dijet MC}} \cdot N_{\mathrm{D}}^{\textrm{dijet MC}}}$ and $R_{\mathrm{B}}^{\mathrm{corr}} = \frac{N_{\mathrm{B}}^{\textrm{dijet MC}} \cdot N_{\mathrm{F}}^{\textrm{dijet MC}}}{N_{\mathrm{C}}^{\textrm{dijet MC}} \cdot N_{\mathrm{E}}^{\textrm{dijet MC}}}$,
\end{center}
where $N^{\textrm{dijet MC}}$ is the number of events predicted by QCD dijet simulation in a given region. The prediction by \textsc{Pythia} is used to correct for this correlation, while the difference between the predictions of the correlation by other QCD dijet simulations (see Section~\ref{sec:datasim}) is used to determine the systematic uncertainty of this 2D sideband method. Experimental systematic uncertainties (see Section~\ref{sec:systunc}) are found to have a negligible impact on $R^{\mathrm{corr}}$. The value of $R^{\mathrm{corr}}$ is found to depend on the signal region and varies between 0.6 at low $m_{tb}$ and 1.3 at high $m_{tb}$. 

The event yields in the different regions considered are shown in Table~\ref{tab:evtyields}. The multi-jet background makes up more than 90$\%$ of the total background in SR1, SR2 and VR, and 75$\%$ of the total background in SR3. The contribution of $t\bar{t}$ events is 4$\%$, 9$\%$ and 25$\%$ of the total background in SR1, SR2 and SR3. The data are well described by the background model. For regions C, D, E and F, the number of multi-jet + $W/Z$+jets is equal to the number of data events after subtracting the $t\bar{t}$ contribution.

\begin{table}[h]
\caption{Event yields in the different regions including the signal regions, SR1, SR2 and SR3. Also shown are the total systematic uncertainties in the estimate of the multi-jet + $W/Z$+jets and $t\bar{t}$ backgrounds in the different regions. The numbers in parentheses are the percentage fractions of the total background. For regions C, D, E and F, the number of multi-jet + $W/Z$+jets is equal to the number of data events after subtracting the $t\bar{t}$ contribution.}
\scalebox{0.68}{%
\begin{tabular}{c|r@{\hskip3pt}c@{\hskip3pt}l|r@{\hskip3pt}c@{\hskip3pt}l|r@{\hskip3pt}c@{\hskip3pt}l|r@{\hskip3pt}c@{\hskip3pt}l|r@{\hskip3pt}c@{\hskip3pt}l|r@{\hskip3pt}c@{\hskip3pt}l}
\hline
\hline
\multicolumn{19}{c}{ \small\textbf{``0 \B-tag in'' category }}  \\
\hline
& \multicolumn{3}{c|}{{\small\textbf{SR1}}}   &  \multicolumn{3}{c|}{{\small\textbf{VR}}} & \multicolumn{3}{c|}{{\small\textbf{Region C}}} & \multicolumn{3}{c|}{{\small\textbf{Region D}}} &\multicolumn{3}{c|}{{\small\textbf{Region E}}} & \multicolumn{3}{c}{{\small\textbf{Region F}}} \\
\hline
{\small Data}   &16,333 && & 57,626 & & &655,669 & & & 267,440  & &    &958,847 & & &12,591,520 & &\\
{\small \ttbar} & 620 &$\pm$& 160 (4$\%$) &780 &$\pm$& 190 (1$\%$) &1,520 &$\pm$& 310 &2,400 &$\pm$& 600   &3,400 &$\pm$& 900   &5,100  &$\pm$& 1,100\\
{\small Multi-jet+$W/Z$+jets} &15,200 & $\pm$ & 2,500 (96$\%$) & 54,000 & $\pm$ & 12,000 (99$\%$)&\multicolumn{3}{c|}{}& \multicolumn{3}{c|}{}& \multicolumn{3}{c|}{}\\
\hline
\multicolumn{19}{c}{ \small\textbf{``1 \B-tag in'' category} } \\
\hline
& \multicolumn{3}{c|}{{\small\textbf{SR3}}}   &  \multicolumn{3}{c|}{{\small\textbf{SR2}}} & \multicolumn{3}{c|}{{\small\textbf{Region C}}} & \multicolumn{3}{c|}{{\small\textbf{Region D}}} &\multicolumn{3}{c|}{{\small\textbf{Region E}}} & \multicolumn{3}{c}{{\small\textbf{Region F}}} \\
\hline
{\small Data}& 4,265 &&& 12,834 & & &78,326 & & &56,044 & & &187,990 & & &1,224,317 & &\\
{\small \ttbar}&1,120 &$\pm$& 290 (25$\%$)&1,140 &$\pm$& 280 (9$\%$) & 1,120 &$\pm$&  260 & 5,300 &$\pm$& 1,200 & 6,700 &$\pm$& 1,500 & 5,600 &$\pm$& 1,200\\
{\small Multi-jet+$W/Z$+jets}&3,250 &$\pm$&970 (75$\%$) &11,200 &$\pm$& 2,700 (91$\%$)& \multicolumn{3}{c|}{} & \multicolumn{3}{c|}{} & \multicolumn{3}{c|}{} \\
\hline
\end{tabular}
}
\label{tab:evtyields}
\end{table}

\section{Systematic uncertainties}
\label{sec:systunc}
The sources of systematic uncertainty can be broadly divided into three groups: those of experimental nature, those related to the modelling in simulation, and those related to the data-driven multi-jet background estimation.

The simulated samples are affected by uncertainties related to the description of the detector response. The dominant detector-related systematic effects are due to the uncertainties in the jet energy scale (JES) and resolution (JER)~\cite{Aaboud:2017jcu}, in the \B-tagging efficiency and mistag rate \cite{ATL-PHYS-PUB-2016-012} and in the top tagging. The main contributions to the uncertainties in the small-$R$ JES, derived as a function of jet \pt\ and $\eta$, are related to in situ calibration, the dependence on the pile-up activity and on the flavour composition of jets~\cite{ATL-PHYS-PUB-2015-015,ATLAS-CONF-2015-002}. The uncertainty in the scale and resolution of large-$R$ jet energy and mass is evaluated by comparing the ratio of calorimeter-based to track-based measurements in multi-jet data and simulation~\cite{PERF-2015-04,ATLAS-CONF-2016-035}. The flavour-tagging efficiency and its uncertainty for $b$-jets~\cite{PERF-2012-04} is estimated in $t\bar{t}$ events, while the misidentification rate for $c$-jets and other jets and their corresponding uncertainties are determined using a $t\bar{t}$-enriched region and multi-jet events, respectively. The SD top-tagging uncertainty is estimated by varying the \pt\ of the subjets used as inputs to the SD algorithm by 2.5$\%$. This value is derived using a procedure described in Ref.~\cite{PERF-2015-04} and is found to cover any data/simulation differences in the $\log \chi_{\textrm{SD}}$ distribution (see Figure~\ref{fig:SD_logchi}). Systematic uncertainties in the lepton veto are found to have a negligible effect.

Flavour-tagging simulation-to-data efficiency correction factors~\cite{PERF-2012-04} depend on the jet \pt\ and $\eta$. These correction factors have several sources of uncertainty. They are split into uncorrelated components that are then treated independently. Additional uncertainties are considered in the extrapolation of the \B-quark jet and $c$-jet efficiency calibration from low \pt, where there is enough data to make a measurement, to high \pt.

The average number of interactions per bunch crossing is rescaled in simulation by 9$\%$ to improve agreement with data, and an uncertainty, as large as the correction, is assigned. Finally, a global normalisation uncertainty of 2.1$\%$ is assigned due to the uncertainty in the luminosity measurement. It is derived, following a methodology similar to that detailed in Ref.~\cite{DAPR-2013-01}, from a calibration of the luminosity scale using x-y beam-separation scans performed in August 2015 and May 2016.

The multi-jet background uncertainties pertain primarily to the estimation method itself. Simulation predictions for the correlation between top- and \B-tagging criteria in the multi-jet background estimation are one source of uncertainty; different event generators are compared to account for differences in modelling of the matrix element and parton showering. The uncertainty in the total background yield arising from the contribution of multi-jets is 15$\%$ in SR1, 22$\%$ in SR2 and 22$\%$ in SR3.

The second largest background, from $t\bar{t}$ events, is assigned a 6$\%$ normalisation uncertainty corresponding to the uncertainty in the production cross-section. An additional uncertainty in the modelling of this background is derived from data/simulation differences observed in the top-quark \pt\ spectrum in $t\bar{t}$ differential cross-section measurements~\cite{TOPQ-2016-01}. This uncertainty has an approximately linear dependence on $m_{tb}$. It is 13$\%$ at $m_{tb}$ = 1 TeV, 31$\%$ at $m_{tb}$ = 2 TeV, 48$\%$ at $m_{tb}$ = 3 TeV and 65$\%$ at $m_{tb}$ = 4~TeV.

The total systematic uncertainty in the background yield is dominated by uncertainties in the 2D sideband method and in the flavour-tagging efficiencies. The dominant systematic uncertainty in the signal yield is due to uncertainties in flavor tagging and top tagging efficiencies. The impact of theoretical uncertainties in the signal acceptance on the results of the analysis is negligible. The statistical uncertainty of the data dominates for $m_{tb} > 2~\TeV$.

\section{Statistical analysis and results}
\label{subsec:ana_stat}
In order to test for the presence of a massive resonance, templates in the variable $m_{tb}$ obtained from the simulated signal event samples and the background events estimated using data-driven methods and simulation, are fit to data, using a binned maximum-likelihood approach based on the RooStats framework~\cite{RooStatsone,RooStatsthree}. The fits are performed simultaneously in the three signal regions (see Section~\ref{subsec:evtsel}). The background processes considered in the maximum-likelihood fit are the dominant multi-jet background and $W/Z$+jets, estimated together using the 2D sideband method (see Section~\ref{subsec:evtsel}), and $t\bar{t}$ events (see Table~\ref{tab:evtyields} for event yields).

The systematic uncertainties described in Section~\ref{sec:systunc} may change the acceptance and shape of the $m_{tb}$ distribution of both a potential \wp-boson signal and the background processes, and are incorporated into the fit as nuisance parameters with log-normal constraints, with correlations across signal regions and signal and background processes taken into account. Some systematic uncertainties, such as those in the JES, affect the shapes of the histogram templates. These systematic uncertainties in the shapes are accounted for by introducing nuisance parameters $\alpha_k$ that describe the possible variation in the shapes of the histograms for each process $k$. A log-normal constraint with mean 0 and width 1 is applied to each of the parameters $\alpha_k$. When performing the maximum-likelihood fit, all of the parameters $\alpha_{k}$ are allowed to vary.

The signal ($s$) and background ($b$) expectations are functions of the nuisance parameters $\vec{\theta}$. These functions are built such that the response of $s$ and $b$ to each $\theta$ is factorised from the nominal value ($s_{0}$) of the expected rate: $s(\vec{\theta}) = s_{0}\times\prod\nu(\vec{\theta})$ where $\nu(\vec{\theta})$ describe the effect of variations in the nuisance parameters $\vec{\theta}$ and similarly for $b$.

The $p$-value $p_{b}$, representing the probability that the data is compatible with the background-only hypothesis, is estimated using the log-likelihood ratio (LLR) test statistic and evaluated using the asymptotic approximation~\cite{Cowan:2010js}. In the absence of any significant excess above the expected background, upper limits at the 95$\%$ CL on the signal production cross-section times branching ratio are derived using the ${CL}_{s}$ method~\cite{CLs}. Limits derived using the asymptotic approximation were cross-checked using pseudo-experiments and found to have a difference of less than 10$\%$ for all \wp\ signal masses.

\begin{figure}[!h!tp]
\centering
\subfloat[]{
\includegraphics[width=0.48\textwidth,trim={0 2cm 0 0},clip]{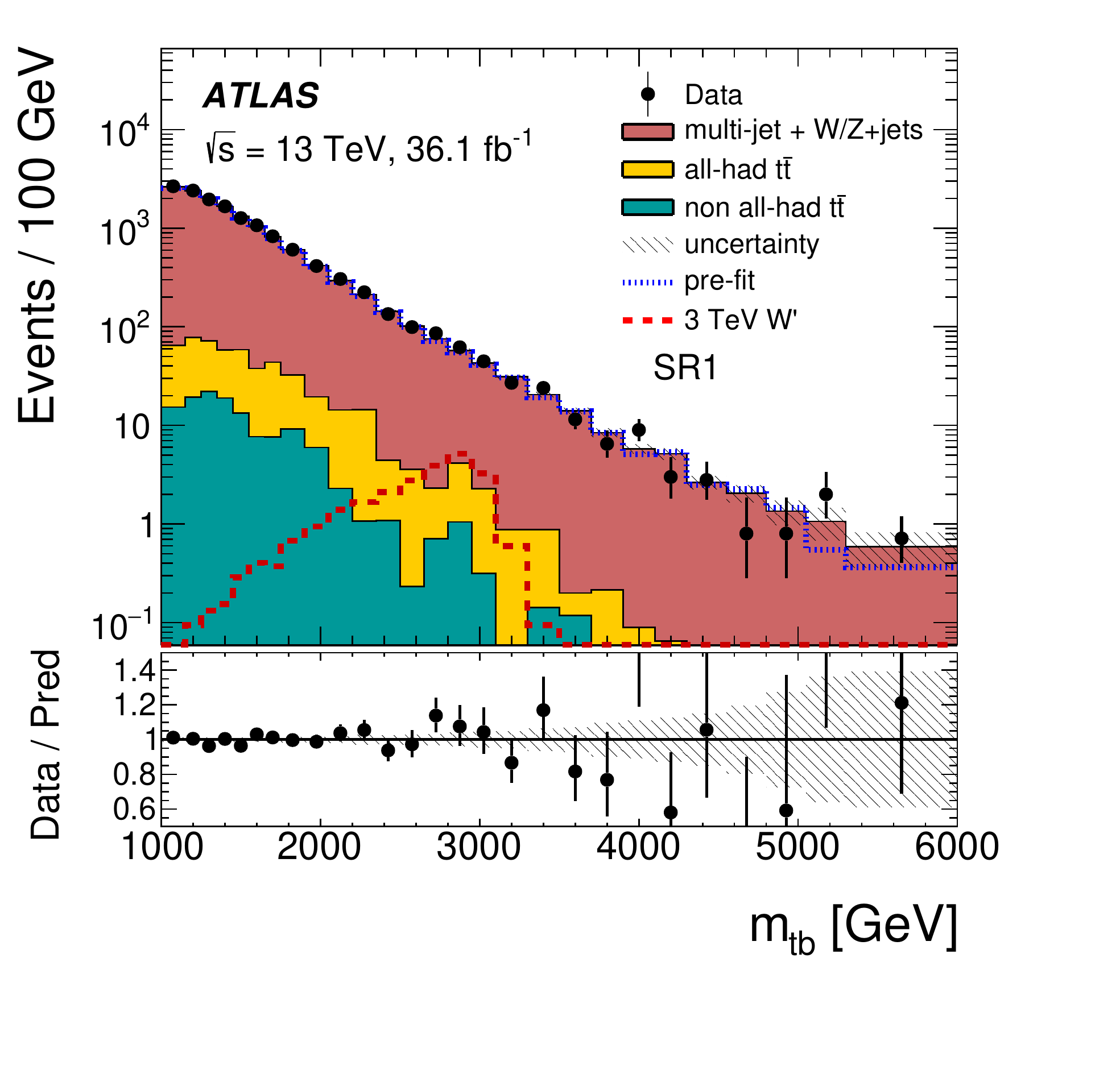}
}
\subfloat[]{
\includegraphics[width=0.48\textwidth,trim={0 2cm 0 0},clip]{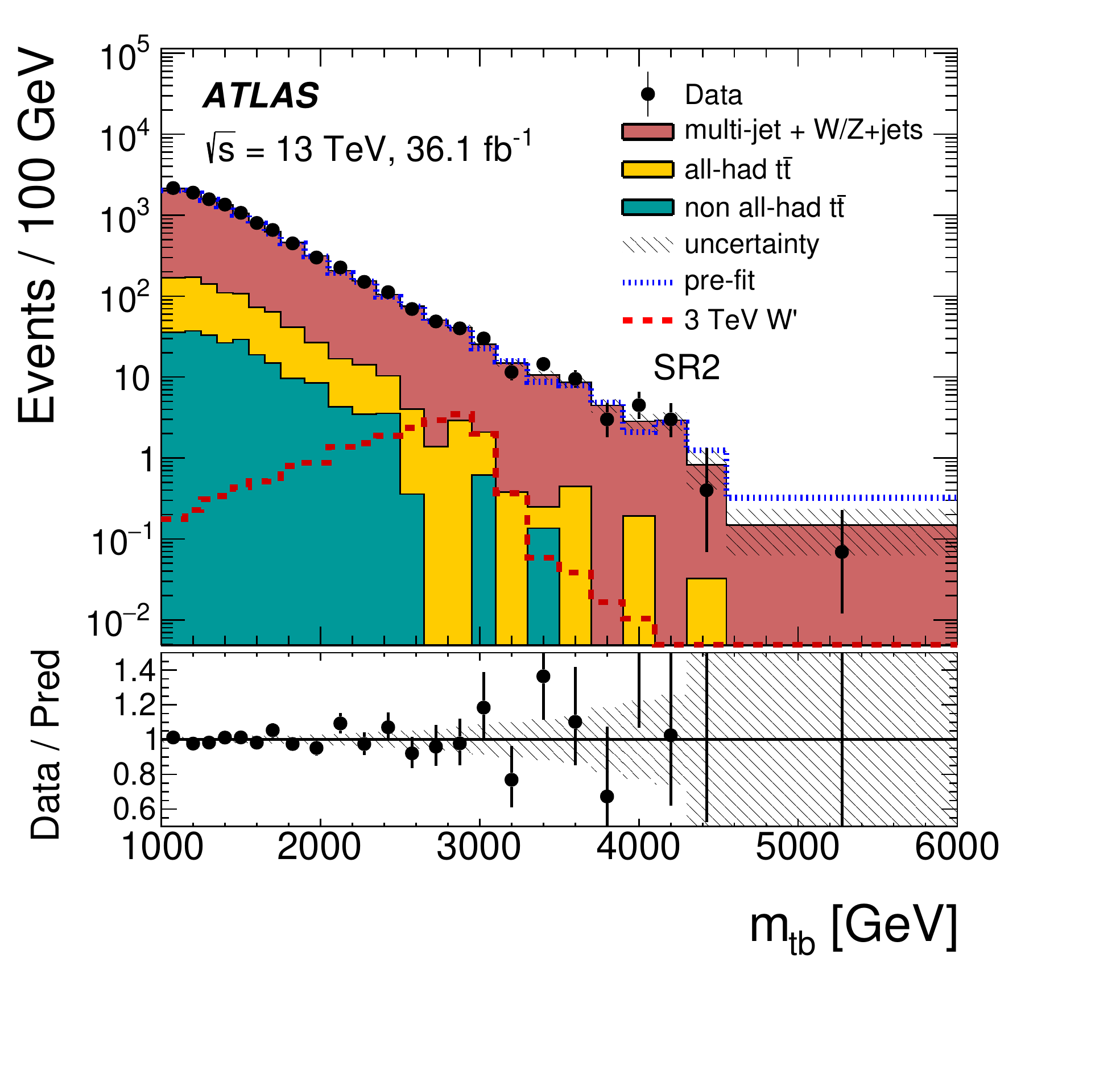}
} \\
\subfloat[]{
\includegraphics[width=0.48\textwidth,trim={0 2cm 0 0},clip]{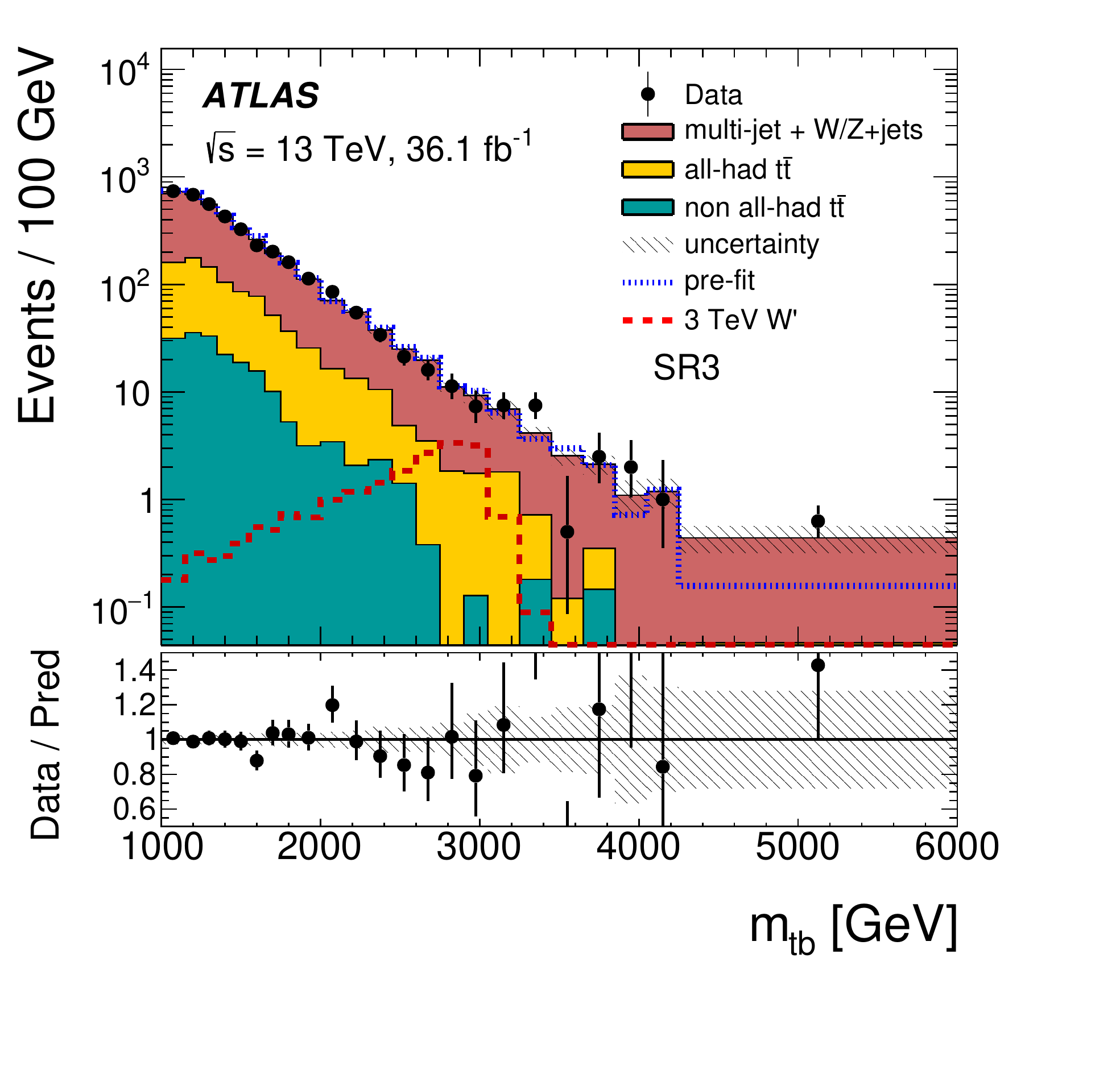}
}
\subfloat[]{
\includegraphics[width=0.48\textwidth,trim={0 2cm 0 0},clip]{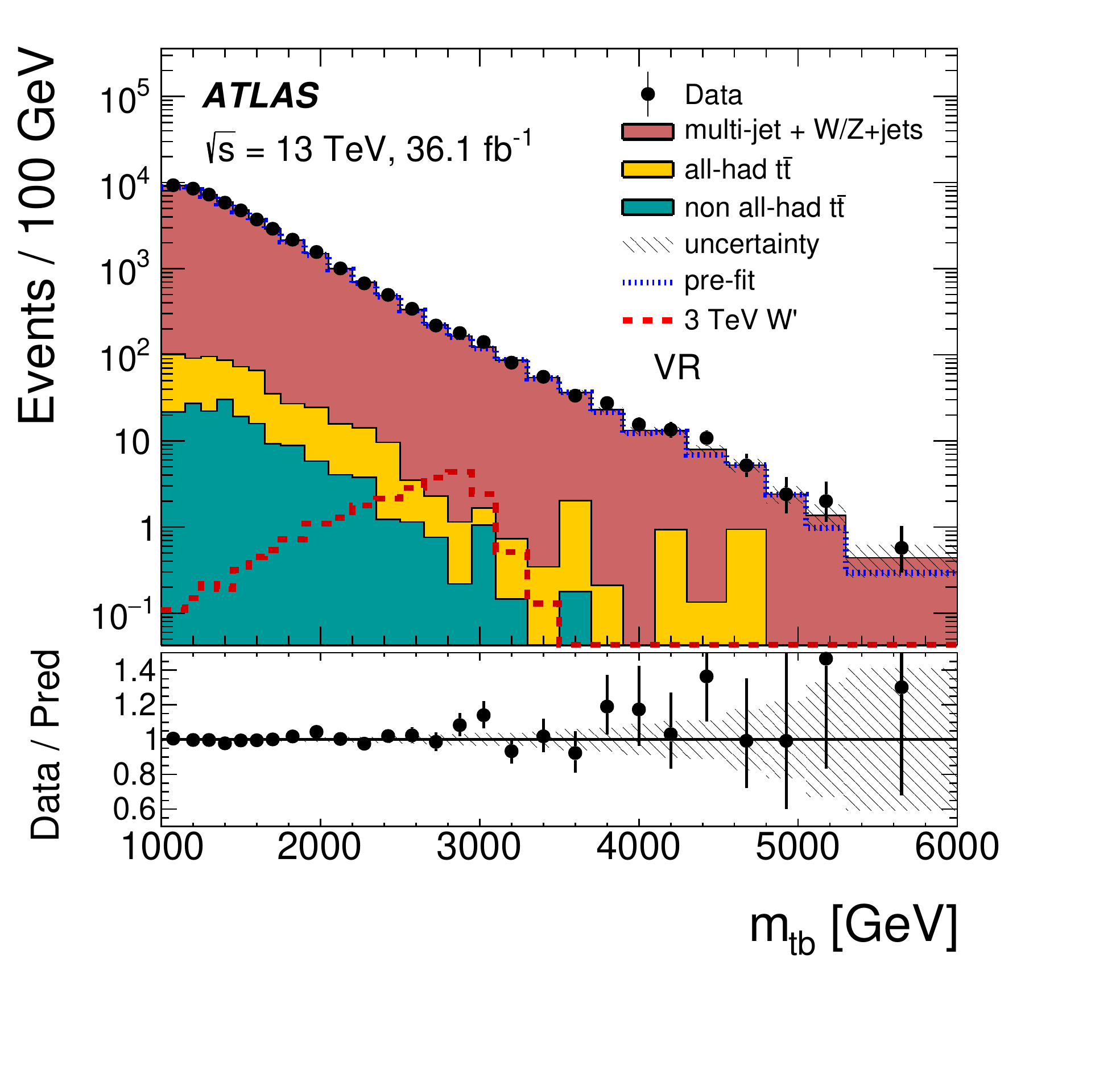}
}
\caption{Reconstructed $m_{tb}$ distributions in data and for the background after the fit to data in the three signal regions and in the multi-jet validation region: (a) SR1, (b) SR2, (c) SR3, and (d) VR. The top panel shows the total-background $m_{tb}$ distribution before the fit to data as the narrow dotted line and the 3 TeV \wpr-boson signal $m_{tb}$ distribution as the dashed line. The ``non all-had $t\bar{t}$~'' label refers to $t\bar{t}$ events in which the $W$ boson from one or both top quarks decays leptonically. The bottom panel of the plot shows the ratio of data to prediction and the hatched band includes the systematic uncertainties after the fit to data.}
\label{fig:inputmtb}
\end{figure}

\begin{figure}[!h!tpb]
  \centering
  \includegraphics[width=0.8\textwidth]{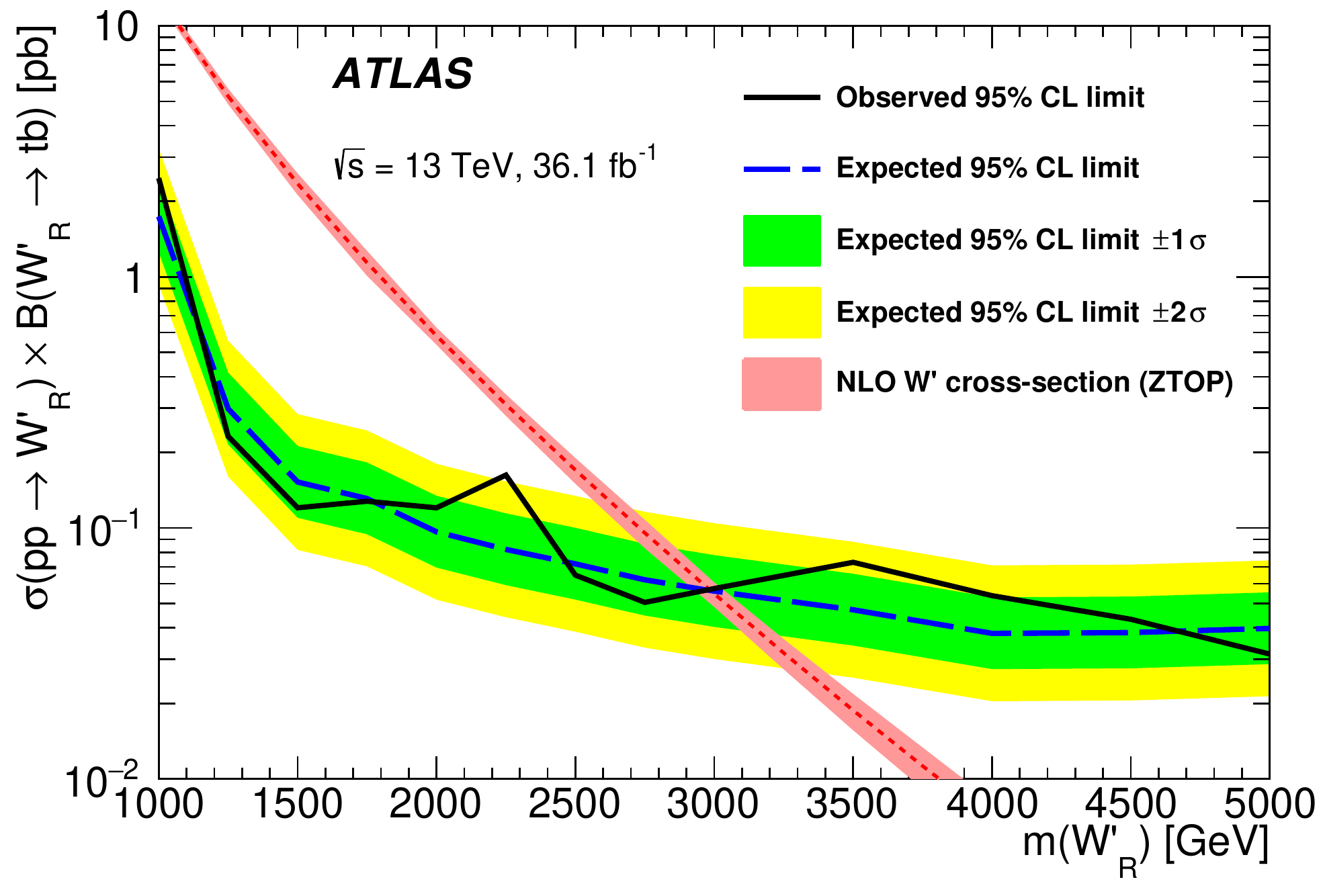}
  \includegraphics[width=0.8\textwidth]{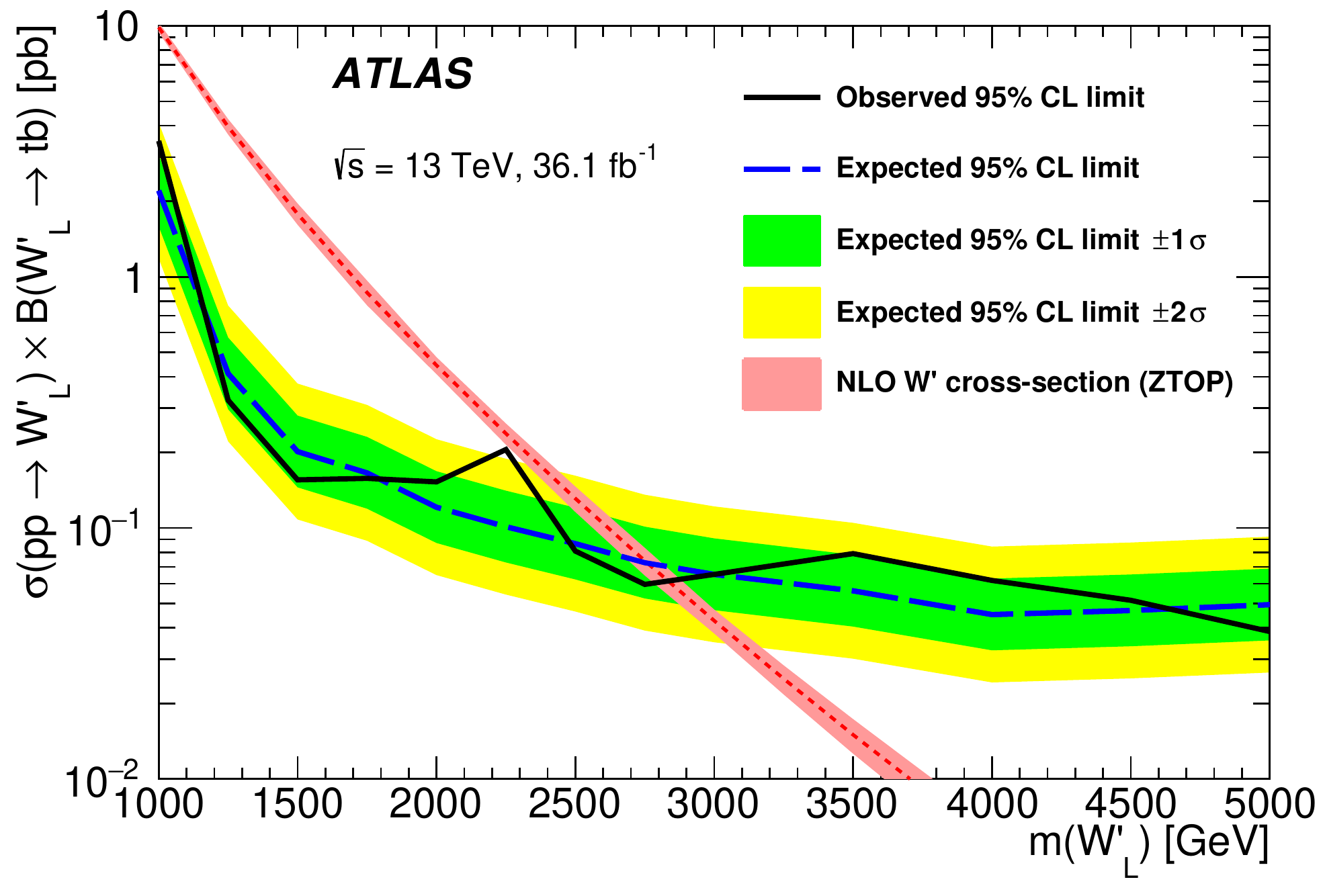}
  \caption{Observed and expected 95\% CL limits on the \wpr-boson (top) and \wpl-boson (bottom) cross-sections times branching ratio of \wp\ to $t\bar{b}$ decay as a function of the corresponding \wp-boson mass. The expected 95$\%$ CL limits are shown with $\pm 1$ and $\pm 2$ standard deviation (s.d.) bands. The narrow dotted curves show the theoretical cross-section predictions and the bands around them show the uncertainties in the predictions for the corresponding \wp-boson signal. The observed and expected limits are derived using a linear interpolation between simulated signal mass hypotheses.}
  \label{fig:stat_limit_asimov}
\end{figure}

Figure~\ref{fig:inputmtb} shows the \mtb\ distributions in the three signal regions and the validation region after the fit to data. The fit in VR is done independently to test the post-fit agreement of the prediction with data. The maximum value of $m_{tb}$ observed in data is 5.8 TeV. The hatched band in the bottom panel includes the systematic uncertainties described in Section~\ref{sec:systunc} after the fit to data. The most discrepant region, at 2.25 TeV, has a local significance of 2.0 $\sigma$ for the combined fit in the three SRs, consistent with the background-only hypothesis. In the absence of any significant excess over the background-only hypothesis, 95$\%$ CL limits are derived on the cross-section times branching ratio of \wp\ to $t\bar{b}$ decay, as shown in Figure~\ref{fig:stat_limit_asimov}, for the right-handed and left-handed couplings. The observed and expected limits are derived using a linear interpolation between simulated signal mass hypotheses. They translate to observed (expected) lower limits on the mass of a \wp\ boson, with the same coupling to fermions as the SM $W$ boson, of 3.0 (3.0) TeV and 2.9 (2.8) TeV in the right- and left-handed models, respectively. These mass limit values are obtained from the intersection of the theory curve and the observed and expected limit curves using a linear interpolation between the 2.75~TeV and 3~TeV \wp-boson signal mass points. The narrow dotted curve in Figure~\ref{fig:stat_limit_asimov} shows the cross-section times branching ratio of \wp\ to $t\bar{b}$ decay calculated with \textsc{Ztop}~\cite{Duffty:2012rf}. The band around this curve shows the uncertainty in the theoretical cross-section obtained by summing in quadrature the uncertainties in the estimation of the parton distribution function, strong coupling constant, renormalization and factorization scales and the top-quark mass. The difference between the mass exclusion limit results for \wpr- and \wpl-boson signals is mainly due to different total cross-sections ($\sigma(pp \rightarrow W') \cdot B(W' \rightarrow t\bar{b})$) of the two processes as discussed in Section~\ref{sec:datasim}.

\FloatBarrier
\section{Summary}
\label{sec:conclusion}
A search for $W' \rightarrow t\bar{b} \rightarrow q\bar{q}' b\bar{b}$ is presented using \lumi\ of $\sqrt{s} = 13~\TeV$ proton--proton collision data collected with the ATLAS detector at the LHC. The analysis makes use of jet substructure tagging optimised to select large-$R$ jets originating from hadronically decaying top quarks using the shower deconstruction algorithm and \B-tagging of small-$R$ jets. The observed \mtb\ spectrum is consistent with the background-only prediction and exclusion limits at 95$\%$ CL are set on the \wp-boson production cross-section times branching ratio to $t\bar{b}$ for right-handed and left-handed couplings as a function of the \wp\ mass in the range 1--5 TeV. Cross-section limits are set at high \wp-boson masses, excluding \wp\ bosons with right-handed couplings with masses below 3.0~\TeV\ and excluding \wp\ bosons with left-handed couplings with masses below 2.9~\TeV (at 95$\%$ CL).

\section*{Acknowledgements}

We thank CERN for the very successful operation of the LHC, as well as the
support staff from our institutions without whom ATLAS could not be
operated efficiently.

We acknowledge the support of ANPCyT, Argentina; YerPhI, Armenia; ARC, Australia; BMWFW and FWF, Austria; ANAS, Azerbaijan; SSTC, Belarus; CNPq and FAPESP, Brazil; NSERC, NRC and CFI, Canada; CERN; CONICYT, Chile; CAS, MOST and NSFC, China; COLCIENCIAS, Colombia; MSMT CR, MPO CR and VSC CR, Czech Republic; DNRF and DNSRC, Denmark; IN2P3-CNRS, CEA-DRF/IRFU, France; SRNSFG, Georgia; BMBF, HGF, and MPG, Germany; GSRT, Greece; RGC, Hong Kong SAR, China; ISF, I-CORE and Benoziyo Center, Israel; INFN, Italy; MEXT and JSPS, Japan; CNRST, Morocco; NWO, Netherlands; RCN, Norway; MNiSW and NCN, Poland; FCT, Portugal; MNE/IFA, Romania; MES of Russia and NRC KI, Russian Federation; JINR; MESTD, Serbia; MSSR, Slovakia; ARRS and MIZ\v{S}, Slovenia; DST/NRF, South Africa; MINECO, Spain; SRC and Wallenberg Foundation, Sweden; SERI, SNSF and Cantons of Bern and Geneva, Switzerland; MOST, Taiwan; TAEK, Turkey; STFC, United Kingdom; DOE and NSF, United States of America. In addition, individual groups and members have received support from BCKDF, the Canada Council, CANARIE, CRC, Compute Canada, FQRNT, and the Ontario Innovation Trust, Canada; EPLANET, ERC, ERDF, FP7, Horizon 2020 and Marie Sk{\l}odowska-Curie Actions, European Union; Investissements d'Avenir Labex and Idex, ANR, R{\'e}gion Auvergne and Fondation Partager le Savoir, France; DFG and AvH Foundation, Germany; Herakleitos, Thales and Aristeia programmes co-financed by EU-ESF and the Greek NSRF; BSF, GIF and Minerva, Israel; BRF, Norway; CERCA Programme Generalitat de Catalunya, Generalitat Valenciana, Spain; the Royal Society and Leverhulme Trust, United Kingdom.

The crucial computing support from all WLCG partners is acknowledged gratefully, in particular from CERN, the ATLAS Tier-1 facilities at TRIUMF (Canada), NDGF (Denmark, Norway, Sweden), CC-IN2P3 (France), KIT/GridKA (Germany), INFN-CNAF (Italy), NL-T1 (Netherlands), PIC (Spain), ASGC (Taiwan), RAL (UK) and BNL (USA), the Tier-2 facilities worldwide and large non-WLCG resource providers. Major contributors of computing resources are listed in Ref.~\cite{ATL-GEN-PUB-2016-002}.

\printbibliography
\clearpage
\begin{flushleft}
{\Large The ATLAS Collaboration}

\bigskip

M.~Aaboud$^\textrm{\scriptsize 137d}$,
G.~Aad$^\textrm{\scriptsize 88}$,
B.~Abbott$^\textrm{\scriptsize 115}$,
O.~Abdinov$^\textrm{\scriptsize 12}$$^{,*}$,
B.~Abeloos$^\textrm{\scriptsize 119}$,
S.H.~Abidi$^\textrm{\scriptsize 161}$,
O.S.~AbouZeid$^\textrm{\scriptsize 139}$,
N.L.~Abraham$^\textrm{\scriptsize 151}$,
H.~Abramowicz$^\textrm{\scriptsize 155}$,
H.~Abreu$^\textrm{\scriptsize 154}$
% ...
\bigskip
\\
% ...
$^{12}$ Institute of Physics, Azerbaijan Academy of Sciences, Baku, Azerbaijan\\
% ...
$^{88}$ CPPM, Aix-Marseille Universit{\'e} and CNRS/IN2P3, Marseille, France\\
% ...
$^{115}$ Homer L. Dodge Department of Physics and Astronomy, University of Oklahoma, Norman OK, United States of America\\
% ...
$^{119}$ LAL, Univ. Paris-Sud, CNRS/IN2P3, Universit{\'e} Paris-Saclay, Orsay, France\\
% ...
$^{137}$ $^{(a)}$ Facult{\'e} des Sciences Ain Chock, R{\'e}seau Universitaire de Physique des Hautes Energies - Universit{\'e} Hassan II, Casablanca; $^{(b)}$ Centre National de l'Energie des Sciences Techniques Nucleaires, Rabat; $^{(c)}$ Facult{\'e} des Sciences Semlalia, Universit{\'e} Cadi Ayyad, LPHEA-Marrakech; $^{(d)}$ Facult{\'e} des Sciences, Universit{\'e} Mohamed Premier and LPTPM, Oujda; $^{(e)}$ Facult{\'e} des sciences, Universit{\'e} Mohammed V, Rabat, Morocco\\
% ...
$^{139}$ Santa Cruz Institute for Particle Physics, University of California Santa Cruz, Santa Cruz CA, United States of America\\
% ...
$^{151}$ Department of Physics and Astronomy, University of Sussex, Brighton, United Kingdom\\
% ...
$^{154}$ Department of Physics, Technion: Israel Institute of Technology, Haifa, Israel\\
$^{155}$ Raymond and Beverly Sackler School of Physics and Astronomy, Tel Aviv University, Tel Aviv, Israel\\
% ...
$^{161}$ Department of Physics, University of Toronto, Toronto ON, Canada\\
% ...
$^{a}$ Also at Department of Physics, King's College London, London, United Kingdom\\
$^{b}$ Also at Institute of Physics, Azerbaijan Academy of Sciences, Baku, Azerbaijan\\
$^{c}$ Also at Novosibirsk State University, Novosibirsk, Russia\\
$^{d}$ Also at TRIUMF, Vancouver BC, Canada\\
$^{e}$ Also at Department of Physics and Astronomy, University of Louisville, Louisville, KY, United States of America\\
% ...
$^{*}$ Deceased
\end{flushleft}


\end{document}